\documentclass[12pt]{article}

\usepackage[margin=1in]{geometry}
\usepackage{amsmath,amssymb,amsthm,amsfonts}
\usepackage{graphicx}
\usepackage{url}
\usepackage{booktabs}
\usepackage{multirow}
\usepackage{algorithm}
\usepackage{algpseudocode}
\usepackage[round,authoryear]{natbib}
\usepackage{setspace}
\usepackage{authblk}
\usepackage{bm}
\usepackage[hidelinks]{hyperref}

\bibpunct{(}{)}{;}{a}{,}{,}

\theoremstyle{plain}
\newtheorem{thm}{Theorem}

\theoremstyle{definition}
\newtheorem{asm}{Assumption}

\allowdisplaybreaks

\title{\bf In-Sample Evaluation of Subgroups Identified by Generic Machine Learning}

\author[1]{Shuoxun Xu\thanks{Email: \texttt{shuoxunxu\_ucb@berkeley.edu}}}
\author[2]{Xinzhou Guo\thanks{Corresponding author. Email: \texttt{xinzhoug@ust.hk}}}
\affil[1]{Department of Biostatistics and Epidemiology, \\
University of California, Berkeley, U.S.A.}
\affil[2]{Department of Mathematics, \\
The Hong Kong University of Science and Technology, Hong Kong SAR, P.R.C.}

\date{\today}

\begin{document}

\maketitle

\begin{abstract}
When a subgroup is identified from the data, it must be evaluated in a replicable way. The usual in-sample approach, which evaluates the post-hoc identified subgroup as predefined, might suffer from selection bias. This issue of in-sample evaluation of data-dependent objects is well recognized but particularly challenging here. Unlike discrete or finite-dimensional data-dependent objects addressed before, the selection bias here is induced by post-hoc identified subgroups, data-dependent sets potentially defined by infinite-dimensional functionals with nonsmooth boundaries known as nonregularity. The out-of-sample approach, which splits data for subgroup identification and evaluation, can help address selection bias but might suffer from efficiency loss and instability. In this paper, we propose a conditional adaptive perturbation approach to remove selection bias in in-sample subgroup evaluation and deliver valid inference on subgroups identified from the whole dataset by generic machine learning, regardless of whether regularity is satisfied. The proposed method is easy-to-compute, allows model-free and even black-box subgroup identification, and achieves full efficiency across broad scenarios of subgroup analysis through a novel theoretical framework of triple robustness linking rates of subgroup identification and nuisance estimation. The merits of the proposed method are demonstrated by a re-analysis of the ACTG 175 trial.
\end{abstract}


\bigskip

\section{Introduction}
\label{s:intro}

In precision medicine, subgroup analysis, the analysis of treatment
effects within subpopulations, is central to uncovering
heterogeneous treatment effects and enabling personalized
interventions.  For example, \citet{fan2017concordance} found that
ZDV+didanosine outperforms ZDV+zalcitabine among ACTG 175 patients
aged over 37.5 years, and the Women's Health Initiative
\citep{rossouw2007postmenopausal} suggested that hormone replacement
therapy's elevated cardiovascular risk in postmenopausal women is
much attenuated in the 50--59 years old subgroup.  Out of 437 clinical
studies in high-impact medical journals, 270 (62\%) reported
subgroup analysis results \citep{gabler2016no}.

How to make replicable statistical evaluation of subgroups is a critical question in subgroup analysis as it informs practitioners the reliability of the subgroup finding. The answer to this question depends on how we define candidate subgroups. When subgroups are predefined prior to examining data, numerous inference procedures, such as fixed designs \citep{mandrekar2009clinical, ziegler2012personalized} and adaptive designs \citep{friede2012conditional, jenkins2011adaptive}, have been proposed for replicable statistical evaluation of subgroups. Although predefined subgroups are widely adopted in practice, post-hoc identified subgroups; i.e. the subgroups defined in data-driven manners, such as generic machine learning, have also gained increasing attentions as in many modern biomedical studies, we lack sufficient prior knowledge and have to rely on data to segment the population \citep{lipkovich2017tutorial}. When subgroups are post-hoc identified, directly applying existing inference procedures for predefined subgroups might be invalid due to selection bias of in-sample evaluation, which arises because subgroup identification and subgroup inference are based on the same data and thus inherently data-dependent \citep{wang2007statistics,luo2023inference}. For example, in the MONET1 study which evaluated motesanib plus carboplatin/paclitaxel in patients with advanced nonsquamous non-small-cell lung cancer, East Asian patients were identified with promising observed treatment effects \citep{kubota2014phase}. However, the follow-up trial failed to replicate the efficacy of the treatment in the East Asian subgroup \citep{kubota2017phase}, and \cite{guo2023robust} argued that such a discrepancy might be partly due to selection bias inherent in post-hoc subgroup inference. Failures to replicate subgroup findings and concerns about selection bias are also noted in other clinical studies, such as the analysis of treatment effects of antiretroviral therapies among HIV-patient subgroups defined by baseline CD4 cell counts, where differential efficacy suggested by post-hoc subgroup analyses of the ACTG 175 trial was not consistently replicated in follow-up studies \citep{carlson2014selection}. As illustrated by the numerical experiments in Example 1 in the Section B of the Supplementary Material, selection bias is evident in post-hoc subgroup inference. This is a critical issue in precision medicine and falls within a well-known challenging task of modern statistical learning: in-sample evaluation of data-dependent objects, where appropriate adjustment of selection bias is needed \citep{nadeau1999inference}.

Selection bias of in-sample evaluation arises and has been intensively studied in several related but different data-dependent scenarios, such as post-selection inference in regression \citep{panigrahi2021integrative} and interactive multiple testing \citep{lei2018adapt}. What make the selection bias of post-hoc subgroup inference unique and particularly challenging is the data-dependent object studied here. Specifically, unlike the discrete or finite-dimensional data-dependent object studied in existing debiased literature, the selection bias here is induced by the post-hoc identified subgroup, a data-dependent set potentially defined by an infinite-dimensional functional with a nonsmooth boundary. In this paper, nonsmooth subgroup boundary means the distribution of individualized treatment effect is not boundedly differentiable near the subgroup boundary, which is also known as nonregularity as detailed in Section \ref{sec:setting}. In subgroup analysis, nonregularity often arises when the covariates used for subgroup identification are discrete, either by nature or due to how they are stored, or there exists a non-negligible subgroup where the individualized treatment effects are homogeneous \citep{laber2014dynamic,shi2020breaking}. In the presence of nonregularity, selection bias is typically more pronounced and difficult to quantify as a small vibration at the boundary caused by data can lead to significant changes to the post-hoc identified subgroup \citep{wang2007statistics}. Besides nonregularity, generic machine learning, which practitioners often use to capture complex treatment effect heterogeneity without imposing strict model assumptions in subgroup identification \citep{liu2019subgroup}, presents another challenge for post-hoc subgroup inference as generic machine learning estimator is typically an infinite-dimensional functional and can be even black-box. When such model-free manners are adopted in subgroup identification, the post-hoc identified subgroup is a data-dependent object defined by an infinite-dimensional functional and selection bias is usually non-negligible because the convergence rate of generic machine learning estimator is typically nonparametric and slower than $n^{-1/2}$ order. Moreover, since the asymptotic distribution of generic machine learning estimators and their functional is not generally available \citep{abadie2008failure,alom2019state}, inference on post-hoc identified subgroups becomes even more complicated when the subgroups are identified by generic machine learning.

To address selection bias of in-sample subgroup evaluation, a widely adopted strategy is out-of-sample evaluation by data split; i.e. conduct subgroup identification in one part of the data and inference in the other part \citep{su2009subgroup,foster2011subgroup}. Splitting the data into two parts, the identified subgroups and the inferential results are naturally independent of each other and thus the selection bias is automatically controlled. However, data split might suffer from power loss as both the identification and the inference of the subgroups only use part of the data, and such power loss reduces the chance of uncovering promising subgroups and is undesirable in subgroup analysis \citep{brookes2001subgroup}. Moreover, by data split, the inference target is indeed the subgroup identified by part of the data, and different splits typically lead to different subsets of data for identification and thus different identified subgroups, which are usually not the same as the subgroups identified from the whole dataset. Such instability issue of out-of-sample evaluation by data split hinders the interpretation of the result and increases the risk of data snooping in subgroup analysis \citep{yu2013stability}.

In this paper, we consider the post-hoc identified subgroups defined as the subpopulation of subjects whose estimated conditional treatment effects or certain utilities based on the whole dataset exceed a given or estimated threshold, a scenario that encompasses most subgroup identification methods \citep{cai2011analysis, chernozhukov2018generic,imai2022statistical}. We develop a conditional adaptive perturbation method to eliminate the selection bias in in-sample subgroup evaluation and provide valid inference on the average treatment effect of the post-hoc identified subgroup when the subgroups are identified by generic machine learning, regardless of whether regularity is satisfied. The proposed method is easy-to-compute and model-free in the sense that subgroup identification can be either parametric or nonparametric, either correctly specified or misspecified, and even black-box. We establish a novel theoretical framework of triple robustness to link the rates of subgroup identification and nuisance function estimation in in-sample subgroup evaluation. The triply robust framework enables us to demonstrate that the proposed method achieves full efficiency as if the subgroup is identified independently of the data, under weak regularity of bounded differentiability and nonparametric rate for generic machine learning in subgroup identification.

Selection bias has been widely recognized as one of the fundamental challenges in post-hoc subgroup inference \citep{guo2023robust,lipkovich2017tutorial}, and some attempts have been made to address this issue. For example, in randomized trials, \cite{zhao2013effectively} tackles the selection bias via perturbation resampling, but the study is built on strong regularity assumptions of twice differentiability and parametric subgroup identification and does not directly aim for inference on post-hoc identified subgroups. To allow certain nonregularity in post-hoc subgroup inference, \cite{guo2021inference} proposes a bootstrap-based debiased inference procedure for the best selected subgroup in randomized trials and \cite{guo2023assessing} extends the method to observational studies, but these methods require parametric rates in subgroup identification and the best selected subgroup is only a special case of the post-hoc identified subgroup considered in this paper. Other subsampling methods, such as repeatedly data split \citep{dusseldorp2014qualitative}, subbagging \citep{shi2020breaking} and $m$-out-of-$n$ bootstrap \citep{chakraborty2010inference}, have been considered for nonregular inference in several related but different scenarios; in particular, inference on the mean outcome of optimal treatment regime when it might not be unique. However, these methods are not directly applicable to post-hoc subgroup inference as illustrated in Example 1 in Supplementary Material~B because here, the evaluation is over the subpopulation instead of the whole population and selection bias induced by subgroup identification needs to be appropriately accounted for. Our protocol is also related to but different from selective inference which typically focuses on parametric or discrete objects, such as selected regression coefficients, instead of subgroups identified by generic machine learning here \citep{lee2016exact,liu2025selective,guglielmini2025selective}. Thus, existing selective inference methods, which often address selection bias via truncated normal approximation or re-estimation of the parametric or discrete object, are not directly applicable for in-sample subgroup evaluation \citep{tian2018selective,fithian2014optimal}. As far as we know, appropriate debiased in-sample methods to infer the subgroups identified from the whole dataset are still lacking especially when nonregularity is allowed and generic machine learning is adopted in subgroup identification, and we aim to bridge the gap in this paper.

In summary, the contributions of the paper lie in both methodology and theory. Methodologically, we propose a debiased inference procedure for the subgroup identified by generic machine learning from the whole dataset when nonregularity is allowed, enabling valid and efficient in-sample evaluation of subgroups in broad practical scenarios. Theoretically, we develop a triply robust condition to link the rate of subgroup identification with nuisance estimation, substantially advancing beyond the doubly robust framework to allow generic machine learning-based subgroup identification in efficient in-sample subgroup evaluation.

The rest of the paper is organized as follows. In Section \ref{sec2}, we state the problem setting and illustrate the selection bias of in-sample subgroup evaluation. In Section \ref{sec3}, we propose the conditional adaptive perturbation method and investigate its theoretical properties. Numerical results including simulation and real data analysis are presented in Sections \ref{sec:simu} and \ref{sec:real} respectively, and Section \ref{sec:disc} concludes the paper by a discussion. The theoretical proofs can be found in the Supplementary Material.

\section{Problem Setting and Challenge}
\label{sec2}

In this section, we mathematically define the post-hoc identified
subgroup based on the whole dataset in Section~\ref{sec:setting} and
illustrate the challenges encountered in in-sample subgroup
evaluation, particularly selection bias when the data-dependent
object is a set defined by an infinite-dimensional functional with a
non-smooth boundary, in Section~\ref{sec:challenge}.

\subsection{Problem Setting: Post-hoc Identified Subgroup}
\label{sec:setting}

Consider a general scenario of observational study of $n$ subjects
with observations $\mathcal{O}_n:=\{\mathbf{O}_i=(Y_i,G_i,\mathbf{Z}_i),i=1,\dots,n\}$,
$n$ i.i.d.\ copies of $\mathbf{O}_0:=(Y_0,G_0,\mathbf{Z}_0)$, where
$\mathbf{Z}_0\in\mathcal{Z}\subset\mathbb{R}^p$ with $p$ possibly larger than $n$,
$G_0\in\{0,1\}$ and $Y_0\in\mathbb{R}$ denote the covariates, the
treatment assignment and the outcome of the patient respectively.
Define the conditional mean outcome function
$h(k,\mathbf{z})=E(Y_0|\mathbf{Z}_0=\mathbf{z},G_0=k)$, the contrasting mean outcome
function $D(\cdot):=h(1,\cdot)-h(0,\cdot)$ and the propensity score
function $\pi(k,\mathbf{z})=P(G_0=k|\mathbf{Z}_0=\mathbf{z})$, where $\mathbf{z}\in\mathcal{Z}$ and
$k\in\{0,1\}$. Under the potential outcome framework, we let $Y_0(k)$
be the potential outcome under treatment $G_0=k$ with $\|Y_0\|_{\infty}<\infty$.
In subgroup analysis literature, $\|Y_0\|_{\infty}<\infty$ is often
assumed to guarantee uniform consistency
\citep{reeve2023optimal,muller2023isotonic} and covers many practical
scenarios, such as testing scores \citep{kobrin2007historical},
survival data \citep{huang2016efficient} and categorical data
\citep{dudley2000detecting}. Throughout the paper, we assume classical
identifiability conditions in causal literature \citep{ding2024first},
particularly the strong ignorability
$(Y_0(0),Y_0(1))\perp G_0|\mathbf{Z}_0$, as stated in details in
Supplementary Material~A.

Under the above setup, we have $D(\mathbf{Z}_0)=E[Y_0(1)-Y_0(0)|\mathbf{Z}_0]$
which is the heterogeneous treatment effect conditional on $\mathbf{Z}_0$.
Assume a larger treatment effect is preferred. Then, a subgroup of
interest is $\{\mathbf{Z}_0:D(\mathbf{Z}_0)\ge c\}$ as it captures the treatment
effect heterogeneity and includes all the subjects whose individualized
treatment effects are above $c$ \citep{bonvini2023minimax}. Here, $c$
can be either a constant or a value estimated from the data, such as
clinical meaningful thresholds or estimated quantiles
\citep{bonvini2023minimax}. Throughout the paper, we focus on a
constant $c$ and details of the estimated $c$ are provided in
Supplementary Material~C.

In practice, since $D(\cdot)$ is usually unknown, most subgroup
identification methods find a generic machine learning estimator
$\widehat{D}(\cdot)$ for $D(\cdot)$ based on the whole dataset
$\mathcal{O}_n$ and consider the post-hoc identified subgroups
$\{\mathbf{Z}_0:\widehat{D}(\mathbf{Z}_0)\ge c\}$. In particular, the best
selected subgroups \citep{guo2021inference} and the tree-based
subgroups \citep{su2009subgroup,loh2015regression} can be viewed as
special cases of the post-hoc identified subgroup considered here;
see \citet{lipkovich2017tutorial} for a review. To inform
practitioners how good the post-hoc identified subgroup is,
appropriate statistical inference on the Post-hoc Identified
Subgroup Average (PISA) treatment effect, which is a data-dependent
quantity,
\begin{equation}
\label{PISA_defn}
PISA(c):=E(Y_0(1)-Y_0(0)|\widehat{D}(\mathbf{Z}_0)\ge c),
\end{equation}
is often desired in drug development and regulatory decision-making
\citep{guo2021inference}. We aim to construct confidence bounds for $PISA(c)$ when
nonregularity and model-free framework are allowed.  Nonregularity means
the subgroup boundary is nonsmooth, i.e.\ the derivative of the
distribution of individualized treatment effect is unbounded near
the boundary, or equivalently Assumption~\ref{A3} is violated.
Model-free means the working model for $\widehat{D}(\cdot)$ can be
parametric or nonparametric, correctly specified or misspecified,
and even black-box where only the predicted values
$\widehat{D}(\mathbf{Z}_i)$ are available.  Under this setup, the
conditional event $\{\mathbf{Z}_0:\widehat{D}(\mathbf{Z}_0)\ge c\}$ is a
data-dependent set potentially defined by an infinite-dimensional
functional with a nonsmooth boundary, substantially different from
the parametric or discrete data-dependent objects of classical
selective inference \citep{lee2016exact}, as detailed in
Section~\ref{sec:challenge}.

Note that when the working model of $\widehat{D}(\cdot)$ is misspecified, $\widehat{D}(\cdot)$ might not be a consistent estimator of $D(\cdot)$. To facilitate our analysis, we assume its limit $\bar{D}(\cdot)$ exists in a
totally bounded and complete functional metric space $(\mathcal{D},\rho)$
with a VC class $\mathcal{D}$ and
$\rho(\widetilde{D}_1,\widetilde{D}_2)=\sup_{\mathbf{Z}_0\in\mathcal{Z}}|\widetilde{D}_1(\mathbf{Z}_0)-\widetilde{D}_2(\mathbf{Z}_0)|$.
This setting covers most generic machine learning working models
used in subgroup identification, including parametric
\citep{zhao2013effectively}, kernel \citep{cai2011analysis},
tree-based \citep{foster2011subgroup}, and deep neural network
\citep{huang2021deep} working models. Moreover, besides estimated heterogeneous treatment effect $\widehat{D}(\cdot)$, the proposed method also
applies to subgroups thresholded by other estimated utilities, such as the multiplicity of estimated treatment effect and safety \citep{luo2023inference}.

\paragraph{Notations.} Let $\widehat{h}$ and $\widehat{\pi}$ be the
estimators for $h$ and $\pi$ respectively. For any index set
$\mathcal{I}\subset[n]$, let $\widehat{h}_{\mathcal{I}}$,
$\widehat{\pi}_{\mathcal{I}}$ and $\widehat{D}_{\mathcal{I}}(\cdot)$
denote the corresponding estimators obtained from subset
$\mathcal{I}$, and $\mathcal{I}^c$ be the complement of $\mathcal{I}$.
The superscript $(-q(i))$ means that, for $q=1,\dots,Q$ evenly sized
data folds, the corresponding estimator is trained based on the data
excluding the fold that the $i$-th observation belongs to.
$\Phi(\cdot)$ denotes the cumulative distribution function of standard
normal. By $a_n=\Theta_p(b_n)$, we mean that there exist constants
$c<C$ such that $P(cb_n<a_n<Cb_n)\to 1$ as $n\to\infty$.

\subsection{Challenge: Selection Bias in Post-hoc Subgroup Inference}
\label{sec:challenge}

In this subsection, we illustrate the selection bias in post-hoc
subgroup inference and explain why the
classical selective inference, out-of-sample evaluation, and related subsampling methods are
not applicable in our setting.

To illustrate the selection bias in post-hoc subgroup inference, we
start with a doubly-robust type identification \citep{bang2005doubly}
of the average treatment effect of the post-hoc identified subgroup
in observational studies,
$PISA(c)=E[\psi(\mathbf{O}_{0},\pi,h)|\widehat{D}(\mathbf{Z}_0)\ge c]$, where
\begin{equation*}
\psi(\mathbf{O}_{0},\pi^*,h^*)=h^*(1,\mathbf{Z}_0)-h^*(0,\mathbf{Z}_0)+\tfrac{G_0}{\pi^*(1,\mathbf{Z}_0)}(Y_0-h^*(1,\mathbf{Z}_0))+\tfrac{1-G_0}{\pi^*(0,\mathbf{Z}_0)}(Y_0-h^*(0,\mathbf{Z}_0))
\end{equation*}
and $\pi^*$ and $h^*$ are any given working models. Treat the
post-hoc identified subgroup as predefined and consider in-sample
subgroup evaluation, a naive sample-mean estimate for $PISA(c)$ is
\begin{equation}
\label{PISA^hat}
\widehat{PISA}(c):=\frac{\sum_{i=1}^{n}\psi(\mathbf{O}_i,\widehat{\pi}^{(-q(i))},\widehat{h}^{(-q(i))})\cdot\mathbb{I}\{\widehat{D}(\mathbf{Z}_i)\ge c\}}{\sum_{i=1}^{n}\mathbb{I}\{\widehat{D}(\mathbf{Z}_i)\ge c\}},
\end{equation}
where the cross-fitting in $\widehat{\pi}^{(-q(i))}$ and
$\widehat{h}^{(-q(i))}$ is a commonly adopted strategy to reduce the
bias induced by the estimation of the outcome and propensity score
models \citep{newey2018cross}. It is clear that $\widehat{PISA}(c)$
suffers from selection bias and might not even follow normal
distribution asymptotically as illustrated in Supplementary
Material~B because $\widehat{D}(\cdot)$ is estimated from the whole
dataset and the in-sample evaluation induces self-correlation in
Eq.~\eqref{PISA^hat}. Here, in-sample evaluation arises because both
the subgroup identification by $\widehat{D}(\cdot)$ and the subgroup
evaluation by $\widehat{PISA}(c)$ are based on the same data
$\mathcal{O}_n$.

The selection bias in $\widehat{PISA}(c)$ is special and particularly complicated
because the post-hoc identified subgroup
$\{\mathbf{Z}_0:\widehat{D}(\mathbf{Z}_0)\ge c\}$ is a data-dependent set
potentially defined by an infinite-dimensional functional with a
nonsmooth boundary. The latter is also known as nonregularity as defined in Section \ref{sec:setting}. In subgroup analysis, one commonly seen scenario of nonregularity is that there exists a point mass at the boundary of the subgroup; i.e. $P(\bar{D}(\mathbf{Z}_0)=c)>0$, which might arise when the covariates $\mathbf{Z}_0$ are discretely distributed or individualized treatment effect $\bar{D}(\cdot)$ is homogeneous in a nonnegligible subgroup. When this happens, a small fluctuation of $\widehat{D}(\cdot)$ can lead to very different post-hoc identified subgroups $\{\mathbf{Z}_0:\widehat{D}(\mathbf{Z}_0)\ge c\}$ and thus, the (standardized) denominator in Eq. \eqref{PISA^hat}, $\sum_{i=1}^{n}\mathbb{I}\{\widehat{D}(\mathbf{Z}_i)\ge c\}/n$, might not even converge, which further deviates $\widehat{PISA}(c)$ from $PISA(c)$. Besides nonregularity, generic machine learning is often encountered in subgroup identification especially when there exists complex relationship between covariates $\mathbf{Z}_0$ and response $Y_0$. In the model-free scenario of subgroup identification, $\widehat{D}(\cdot)$ is typically a generic machine learning estimator of infinite dimensions, which can be even black-box, and $\{\mathbf{Z}_0:\widehat{D}(\mathbf{Z}_0)\ge c\}$ is thus a data-dependent set defined by infinite dimensional functional $\widehat{D}(\cdot)$. Even though $\sum_{i=1}^{n}\mathbb{I}\{\widehat{D}(\mathbf{Z}_i)\ge c\}/n$ might converge under certain regularity, the convergence rate can be slower than $n^{-1/2}$ due to the potentially nonparametric rate of generic machine learning estimator $\widehat{D}(\cdot)$, which makes the selection bias nonnegligible under the usual $n^{-1/2}$ scale. Moreover, the asymptotic distribution of $\widehat{PISA}(c)$ might be untractable as $\widehat{PISA}(c)$ is a functional of $\widehat{D}(\cdot)$ and the asymptotic distribution of a functional of generic machine learning estimator is not generally available \citep{hirano2012impossibility} and, particularly, might not be normal as illustrated in Supplementary Material~B. Classical selective inference
\citep{lee2016exact,tian2018selective}, which typically conditions
on parametric or discrete data-dependent objects and quantifies
uncertainty via re-estimation or truncated normal approximation, is
not applicable here because the conditional event $\{\mathbf{Z}_0:\widehat{D}(\mathbf{Z}_0)\ge c\}$ is a data-dependent set without
parametric representations or generally available distributions.

To address selection bias in in-sample subgroup evaluation, a natural alternative is out-of-sample evaluation: split the data
into $\mathcal{I}$ and $\mathcal{I}^c$, identify
$\{\mathbf{Z}_0:\widehat{D}_{\mathcal{I}}(\mathbf{Z}_0)\ge c\}$ on
$\mathcal{I}$, and evaluate it on $\mathcal{I}^c$.  But the target
then becomes
$PISA_{\mathcal{I}}(c)=E(Y_0(1)-Y_0(0)|\widehat{D}_{\mathcal{I}}(\mathbf{Z}_0)\ge c)$,
which generally differs from $PISA(c)$ because
$\widehat{D}_{\mathcal{I}}(\cdot)$ uses only part of the data and
changes with the split, so data split both loses power
\citep{sun2010subgroup} and is unstable, inviting selective
reporting through repeated splits.  This mismatch between
subgroups defined by part of the data and those defined by the
whole dataset also invalidates existing subsampling schemes for
nonregular inference over the whole population, including
repeated data split \citep{zhao2013effectively}, subbagging
\citep{shi2020breaking}, and $m$-out-of-$n$ bootstrap
\citep{bickel2012resampling}, as illustrated in Example~1 of
Supplementary Material~B.

\section{Methodology and Theory}
\label{sec3}
In this section, we propose a conditional adaptive perturbation method for valid post-hoc subgroup inference  in Section \ref{sec:method}, and explore the theoretical properties of the proposed method, particularly its full efficiency through a novel theoretical framework of triple robustness linking rates of subgroup identification with nuisance estimation in Section \ref{sec:theory}.

\subsection{Proposed Method: Conditional Adaptive Perturbation}
\label{sec:method}

To address the above stated issues in
Section~\ref{sec:challenge} and conduct valid inference on the
post-hoc identified subgroups, we propose a conditional adaptive
perturbation approach and construct a confidence bound for $PISA(c)$.
The key idea of the proposed method is to (1) adaptively perturb the
dataset and (2) keep $\widehat{D}(\cdot)$ fixed to address selection
bias and stabilize distribution in in-sample evaluation of subgroups
identified by generic machine learning. Details are summarized in
Algorithm~\ref{alg-con-moon}.

\begin{algorithm}[t]
\caption{Confidence bound for $PISA(c)$}
\label{alg-con-moon}
\begin{algorithmic}[1]
\Require Dataset $\mathcal{O}_n$, adaptive perturbation size $m\leq n$,
Monte Carlo repetitions $M$, significance level $\alpha$.
\State Compute the pivotal statistic
$\widehat{PISA}_{\mathcal{I}}(c)=\dfrac{\sum_{i\in\mathcal{I}}\psi(\mathbf{O}_i,\widehat{\pi}^{(-q(i))},\widehat{h}^{(-q(i))})I\{\widehat{D}(\mathbf{Z}_i)\ge c\}}{\sum_{i\in\mathcal{I}}I\{\widehat{D}(\mathbf{Z}_i)\ge c\}}$,
where $\mathcal{I}\subseteq[n]$, $|\mathcal{I}|=m$.
\For{$j=1,\ldots,M$} \Comment{conditional adaptive perturbation}
    \State Generate $m$ independent random variables $V_i\sim N(1,1)$,
    $i=1,\ldots,m$.
    \State Compute
    $\widehat{PISA}_{\mathcal{I}}^{*(j)}(c)\leftarrow
    \dfrac{\sum_{i\in\mathcal{I}}V_i\psi(\mathbf{O}_i,\widehat{\pi}^{(-q(i))},\widehat{h}^{(-q(i))})I\{\widehat{D}(\mathbf{Z}_i)\ge c\}}{\sum_{i\in\mathcal{I}}V_iI\{\widehat{D}(\mathbf{Z}_i)\ge c\}}$.
\EndFor
\State Compute empirical quantiles $c_{\alpha/2,\mathcal{I}}^*$ and
$c_{1-\alpha/2,\mathcal{I}}^*$ of
$\{m^{1/2}(\widehat{PISA}_{\mathcal{I}}^{*(j)}(c)-\widehat{PISA}_{\mathcal{I}}(c))\}_{j=1}^{M}$.
\State \Return
$[\widehat{PISA}_{\mathcal{I}}(c)-c_{\alpha/2,\mathcal{I}}^*m^{-1/2},\widehat{PISA}_{\mathcal{I}}(c)-c_{1-\alpha/2,\mathcal{I}}^*m^{-1/2}]$.
\end{algorithmic}
\end{algorithm}

Algorithm~\ref{alg-con-moon} approximates the distribution of the
pivotal statistic $\widehat{PISA}_{\mathcal{I}}(c)$ by a perturbation
procedure conditional on $\widehat{D}(\cdot)$ fixed and restricted
to a subset $\mathcal{I}$ of adaptive size $m$ (selected by a
data-adaptive procedure in Supplementary Material~E).  Under
regularity and a triply robust condition allowing nonparametric
rates for subgroup identification, the method achieves $\sqrt{n}$
full efficiency as if $\widehat{D}(\cdot)$ were predefined
(Section~\ref{sec:theory}).  The required regularity is weaker
than twice differentiability used for $\sqrt{n}$-scale post-hoc
inference elsewhere \citep{zhao2013effectively,zhang2012robust},
and the triply robust rate condition is weaker than the parametric
rate usually demanded for such inference and covers many practical
generic machine learning-based subgroup identification scenarios.

\subsection{Theory: Full Efficiency and Triple Robustness}
\label{sec:theory}

To demonstrate the full efficiency and validity of the proposed method, we make a classical doubly robust rate requirement in Assumption \ref{A4}. Then,  the first part of Theorem~\ref{thm1}
states that the proposed method achieves full efficiency as if
$\widehat{D}(\cdot)$ is identified independently of the data with
$m=n$, when regularity and a triply robust condition allowing
nonparametric rate of subgroup identification are satisfied as stated
in Assumptions~\ref{A3} and~\ref{A5} respectively. These two
assumptions are mild and cover many practical scenarios of in-sample
subgroup evaluation as detailed later. In addition, regardless of
whether regularity is satisfied and what generic machine learning
model is adopted in subgroup identification, the second part of
Theorem~\ref{thm1} shows that the proposed method delivers valid
inference on $PISA(c)$ when the adaptive perturbation size $m$ is of
smaller order than $n$, ensuring the safety of the proposed method.
In practice, $m$ might be chosen by a data-adaptive method, Algorithm 2, as detailed in  Supplementary Material~E, where Proposition~D.1 shows that the tuning helps the proposed method attain full efficiency by selecting $m=n$ almost surely when both Assumptions~\ref{A3} and~\ref{A5} hold, and ensures its safety by selecting $m=o(n)$ when either is violated.

\begin{asm}[Doubly Robust Rate Requirements]
\label{A4}
The convergence rates of $\widehat{\pi}$ and $\widehat{h}$ satisfy:
$\sup_{\mathbf{Z}_0\in\mathcal{Z}}|\widehat{h}^{(-q(i))}(k,\mathbf{Z}_0)-h(k,\mathbf{Z}_0)|,\sup_{\mathbf{Z}_0\in\mathcal{Z}}|\widehat{\pi}^{(-q(i))}(\mathbf{Z}_0)-\pi(\mathbf{Z}_0)|\rightarrow_p 0$
and
\begin{align*}
E\bigl[(\widehat{h}^{(-q(i))}(k,\mathbf{Z}_i)-h(k,\mathbf{Z}_i))^2\bigr]\cdot E\bigl[(\widehat{\pi}^{(-q(i))}(\mathbf{Z}_i)-\pi(\mathbf{Z}_i))^2\bigr]=o_p(n^{-1}).
\end{align*}
\end{asm}

\begin{asm}[Regularity]
\label{A3}
The maximum local derivative around $\bar{D}(\cdot)$ and $c$ satisfies
$\sup_{\widetilde{D}\in\mathcal{D}\land\rho(\widetilde{D},\bar{D})\leq\delta\land c'\in[c-\delta,c+\delta]}\left|dP(\widetilde{D}(\mathbf{Z}_0)\ge c')/dc'\right|<\infty$,
where $\delta>0$ is a constant.
\end{asm}

\begin{asm}[Triply Robust Condition for Subgroup Identification]
\label{A5}
There exists $d_n$ satisfying $P\left(\sup_{\mathbf{Z}_0}|\widehat{D}(\mathbf{Z}_0)-\bar{D}(\mathbf{Z}_0)|\ge d_n\right)=o(1)$,
where $d_n\cdot E\left[\left(\widehat{h}^{(-q(i))}(k,\mathbf{Z}_i)-h(k,\mathbf{Z}_i)\right)^2\right]$
and $d_n\cdot E\left[\left(\widehat{\pi}^{(-q(i))}(\mathbf{Z}_i)-\pi(\mathbf{Z}_i)\right)^2\right]=o_p(n^{-1})$.
\end{asm}

\begin{thm}
\label{thm1}
Under Assumption~\ref{A4} and as $n\to\infty$, the following equation
holds
\begin{equation*}
\label{Valid-infer}
P\left(c_{1-\alpha/2,\mathcal{I}}^*\leq\sqrt{m}(\widehat{PISA}_{\mathcal{I}}(c)-PISA(c))\leq c_{\alpha/2,\mathcal{I}}^*\right)\to 1-\alpha
\end{equation*}
(1) (Full Efficiency) when $m=n$ and achieves full efficiency as if
$\widehat{D}(\cdot)$ is predefined under Assumptions~\ref{A3}
and~\ref{A5}, or (2) (Safety) when $m=o(n)$.
\end{thm}

To demonstrate full efficiency of the proposed method, we make two mild assumptions, Assumptions \ref{A3} and \ref{A5}, in the first part of Theorem \ref{thm1}. Assumption \ref{A3} requires that within a neighborhood of $\bar{D}(\cdot)$, the derivative of the distribution is bounded around the boundary at $c$, which is used to characterize regularity in this paper and weaker than the regularity assumption used for $\sqrt{n}$-scale post-hoc inference in the existing literature, such as twice differentiability in \cite{zhao2013effectively}. Assumption \ref{A5} imposes a requirement on the convergence rate of generic machine learning estimator $\widehat{D}(\cdot)$, which is stricter when the convergence rates of $\widehat{h}$ and $\widehat{\pi}$ in Assumption \ref{A4} are slower and vice versa. Thus, we call Assumption \ref{A5} triply robust rate for subgroup identification. The joint rate requirement by Assumptions \ref{A4} and \ref{A5} is weaker than the parametric assumption and can be satisfied in broad scenarios of generic machine learning-based subgroup identification allowing nonparametric rates of all the working models. Examples can be found in Supplementary Material~D.

The triple robustness in Assumption \ref{A5} is a nontrivial advancement beyond the double robustness in causal inference \citep{bang2005doubly} to link the convergence rate of the subgroup identification estimator $\widehat{D}(\cdot)$ with those of the nuisance functions $\widehat{h}(\cdot)$ and $\widehat{\pi}(\cdot)$ in subgroup analysis. This novel framework enables us to establish full efficiency of the proposed method across broad scenarios of generic machine learning-based subgroup identification with nonparametric rates, which can be even black-box. Specifically, the proposed method can achieve full efficiency as long as any two of the three components, subgroup identification, outcome regression and propensity score, converge in sufficiently fast nonparametric rates as detailed in Assumption \ref{A5}. Without linking the subgroup identification rate with the others, previous theoretical frameworks of subgroup evaluation typically accommodate parametric subgroup identification only \citep{zhao2013effectively}. In summary, the framework of triple robustness in Assumption \ref{A5} paves a novel way to incorporate the rate of subgroup identification in subgroup evaluation and provides an appropriate theoretical foundation of efficient in-sample subgroup evaluation when generic machine learning is adopted in subgroup identification.

\section{Simulations}
\label{sec:simu}

In this section, we study the finite sample performance of the
proposed method via Monte Carlo simulation. Let the sample size
$n=1000$ and generate the treatment indicator $G_0$ by
$\operatorname{Bernoulli}((1+e^{-0.5(Z_0^{(1)}+Z_0^{(2)})})^{-1})$ and
the response $Y_0$ by
$Y_0=h(0,\mathbf{Z}_0)+G_0\cdot D(\mathbf{Z}_0)+e_0$,
where $\mathbf{Z}_0=(Z_0^{(1)},Z_0^{(2)})^{\top}$ and $e_0\sim N(0,0.4^2)$.
We consider both the regular and nonregular case by varying
$D(\cdot)$ and the generating mechanism of $\mathbf{Z}_0$ as summarized in
settings (A)--(D) in Table~\ref{tab1}. In particular, for settings
(C) and (D), we consider a mixture distribution of $\mathbf{Z}_0$ as
follows,
$Z_0^{(2)}=TU+(1-T)X$
with $T\sim\operatorname{Bernoulli}(0.5)$, $U\sim\operatorname{U}[-1,1]$, and
$P(X=-1)=P(X=-0.8)=P(X=0.8)=P(X=1)=\tfrac{1}{16}$, $P(X=0)=\tfrac{3}{4}$,
which has a point mass. Thus, settings (C) and (D) are nonregular
while settings (A) and (B) are regular. We also consider different
generic machine learning working models for $D(\cdot)$ with settings
(A) and (C) parametric and settings (B) and (D) nonparametric.
Throughout this section, we estimate $\pi$ and $h$ by logistic
regression and B-spline respectively.

\begin{table}[htbp]
\caption{Simulation settings where ``parametric'' refers to linear
$\widehat{D}(\cdot)$ and ``nonparametric'' refers to B-spline
$\widehat{D}(\cdot)$, and $F_{\mathbf{Z}_0}$ is the distribution function
for each dimension of $\mathbf{Z}_0$.}
\label{tab1}
\centering
\footnotesize
\begin{tabular}{lllll}
\toprule
Setting & $h(0,\mathbf{Z}_0)$ & $D(\mathbf{Z}_0)$ & $F_{\mathbf{Z}_0}$ & $\widehat{D}(\cdot)$ \\
\midrule
(A) & $\mathbb{I}\{Z_0^{(2)}\ge 0\}-0.06$ & $\mathbb{I}\{Z_0^{(2)}\ge 0\}-0.06$ & $(\operatorname{U}[-1,1],\operatorname{U}[-1,1])$ & parametric \\
(B) & $\mathbb{I}\{Z_0^{(2)}\ge 0\}-0.06$ & $\mathbb{I}\{Z_0^{(2)}\ge 0\}-0.06$ & $(\operatorname{U}[-1,1],\operatorname{U}[-1,1])$ & nonparametric \\
(C) & $\mathbb{I}\{Z_0^{(2)}\ge 0\}\cdot|Z_0^{(2)}|^{1/3}$ & $[Z_0^{(2)}-0.95\operatorname{sign}(Z_0^{(2)})]\mathbb{I}\{|Z_0^{(2)}|\leq 0.95\}$ & $(\operatorname{U}[-1,1],\text{Mixture})$ & parametric \\
(D) & $\mathbb{I}\{Z_0^{(2)}\ge 0\}\cdot|Z_0^{(2)}|^{1/3}$ & $[Z_0^{(2)}-0.95\operatorname{sign}(Z_0^{(2)})]\mathbb{I}\{|Z_0^{(2)}|\leq 0.95\}$ & $(\operatorname{U}[-1,1],\text{Mixture})$ & nonparametric \\
\bottomrule
\end{tabular}
\end{table}

We set $M=2000$ perturbations and 500 repetitions, and compare:
(1) the naive in-sample approach by Eq.~\eqref{PISA^hat} using
normal approximation; (2) Sample Split (SS), which uses half the
data for identification and half for evaluation; (3) Oracle, which
treats $\widehat{D}(\cdot)$ as known; and (4) the proposed method
with $m=n,n/2,n/4,n/8$ and $m$ chosen by the procedure of
Section~E of the Supplementary Material with $C=1/2$.  Table~\ref{tab2} reports
Empirical Coverage Probability (ECP) and average Confidence
Interval Length (CIL) for two-sided 95\% intervals of $PISA(c)$.

\begin{table}[htbp]
\caption{Empirical Coverage Probability (ECP) and average Confidence
Interval Length (CIL) of the two-sided 95\% confidence intervals for
$PISA(c)$. The standard error of ECP under the nominal level is
below 0.01.}
\label{tab2}
\centering
\footnotesize
\begin{tabular}{llrrrrrrrr}
\toprule
 & & & & & \multicolumn{5}{c}{Proposed} \\
\cmidrule(lr){6-10}
Setting & & Naive & SS & Oracle & $m=n$ & $m=n/2$ & $m=n/4$ & $m=n/8$ & Alg.~2 \\
\midrule
(A) & ECP & 91.2 & 74.6 & 94.6 & 93.4 & 94.8 & 95.4 & 94.4 & 93.2 \\
    & CIL & 0.15 & 0.22 & 0.17 & 0.15 & 0.22 & 0.31 & 0.43 & 0.16 \\
(B) & ECP & 91.0 & 69.6 & 93.4 & 93.4 & 95.2 & 94.0 & 94.2 & 93.6 \\
    & CIL & 0.17 & 0.23 & 0.14 & 0.17 & 0.23 & 0.33 & 0.47 & 0.17 \\
(C) & ECP & 88.5 & 78.7 & 94.8 & 86.9 & 89.4 & 92.2 & 94.2 & 94.2 \\
    & CIL & 0.31 & 0.44 & 0.29 & 0.17 & 0.24 & 0.34 & 0.48 & 0.44 \\
(D) & ECP & 78.2 & 57.1 & 96.6 & 80.8 & 89.8 & 93.4 & 93.2 & 93.8 \\
    & CIL & 0.28 & 0.35 & 0.26 & 0.16 & 0.22 & 0.32 & 0.47 & 0.42 \\
\bottomrule
\end{tabular}
\end{table}

Table~\ref{tab2} shows that the naive method undercovers in all
settings because of selection bias, especially under nonregularity
and nonparametric identification, and the sample split method
undercovers too since it targets $PISA_{\mathcal{I}}(c)$ rather than
$PISA(c)$.  The proposed method covers well when $m$ is chosen
appropriately: in the regular settings (A) and (B) any $m$
preserves coverage, with larger $m$ giving shorter intervals; in
the nonregular settings (C) and (D), a smaller $m$ is needed to
preserve coverage, matching our theory.  With the data-adaptive
$m$, the proposed method is close to the nominal level across all
settings, and the interval length approaches the oracle in the regular setting (A)
and (B).

\section{Application: The ACTG 175 Trial}
\label{sec:real}

In this section, we apply the proposed method to the AIDS Clinical
Trials Group Protocol 175 (the ACTG 175 trial).  While designed for
observational studies, the method applies here since a randomized
trial is a special case with randomized treatment assignment, and
doubly-robust inference is widely used in trials for efficiency improvement
\citep{zhao2015doubly}.  The trial compares zidovudine (ZDV) and
didanosine (ddI) monotherapies with the combination therapies
ZDV+ddI and ZDV+zalcitabine (zal), in HIV patients with CD4 T cell
counts between 200 and 500 per cubic millimeter
\citep{hammer1996trial}.  Antiretroviral therapies (ART) are known
to exhibit treatment-effect heterogeneity, and subgroup analysis
is commonly conducted to inform treatment assignment
\citep{rothwell1995can}; for example, \citet{fan2017concordance}
identified that patients aged under 35 with higher baseline CD4
counts ($>$300 cells/mm$^3$) might respond better to combination
therapy than to monotherapy.  Such analyses typically ignore
selection bias, and prior work suggests that nonregularity may be
present in the ACTG 175 data when generic machine learning (e.g.\
smoothing spline) is used for identification
\citep{fan2017concordance}, motivating a debiased post-hoc
analysis.

\begin{figure}[htbp]
\centering
\includegraphics[width=0.45\textwidth]{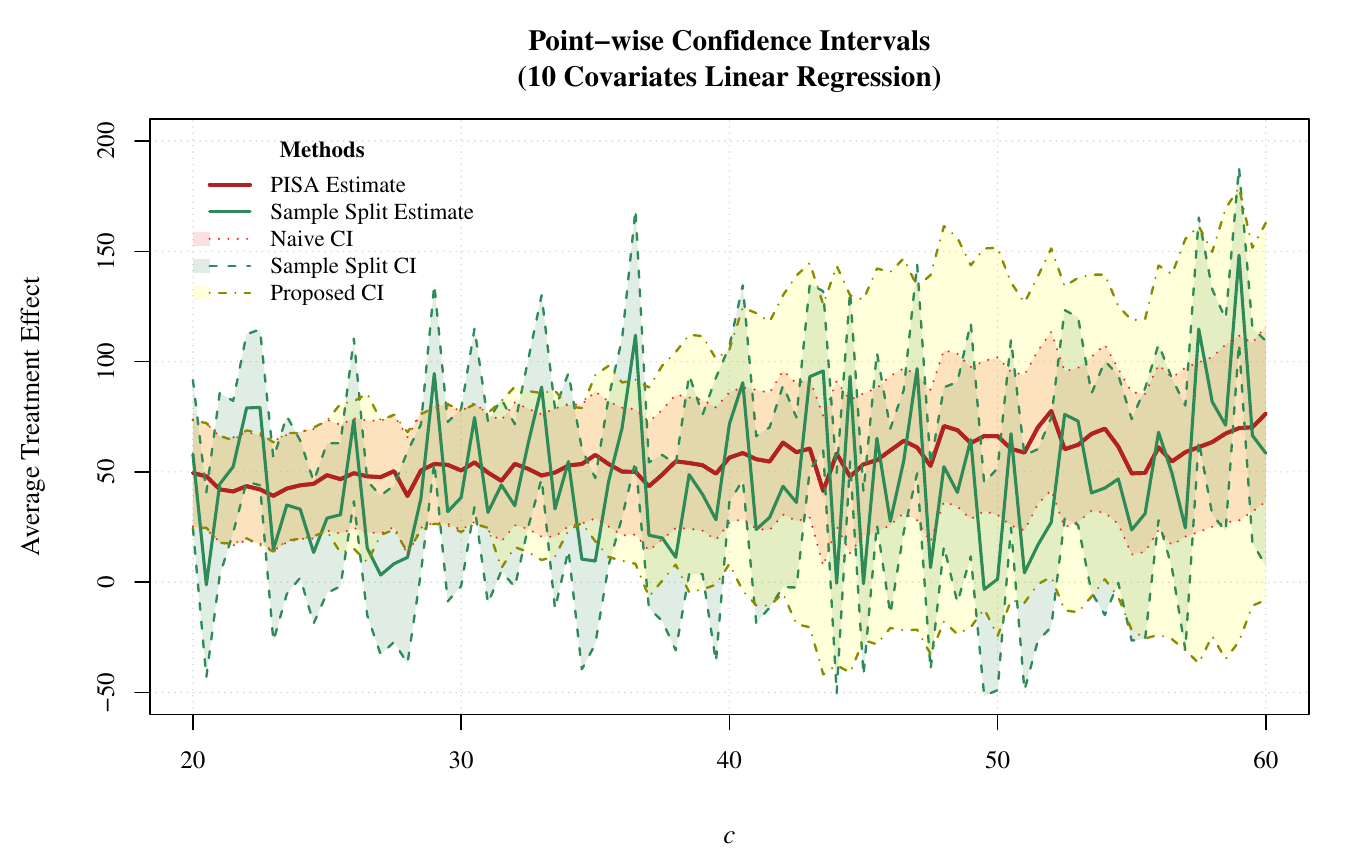}\hspace{0.05\textwidth}%
\includegraphics[width=0.45\textwidth]{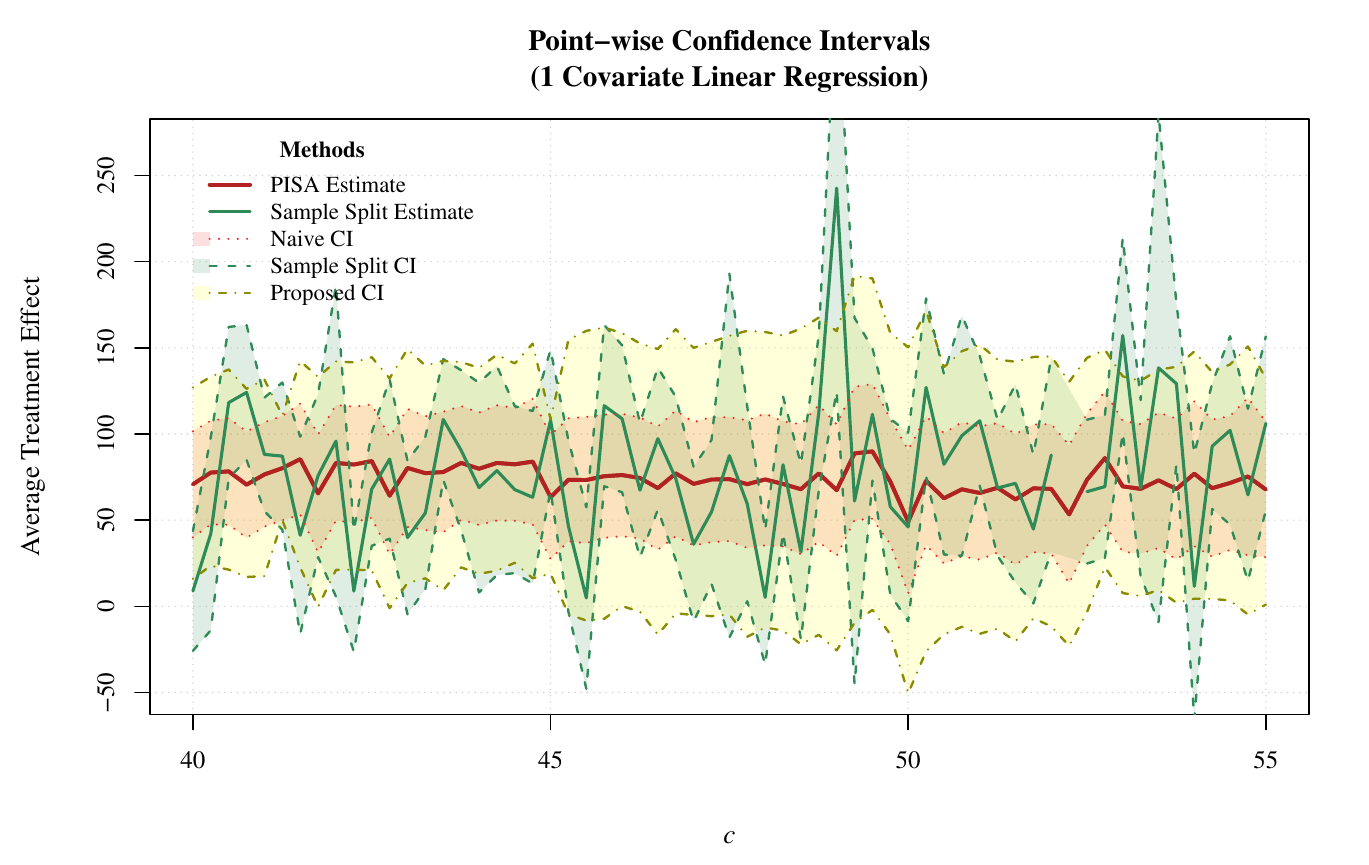}
\caption{Pointwise 95\% confidence intervals for $PISA(c)$. Left
panel: 10-covariate simple linear regression for
$\widehat{D}(\cdot)$. Right panel: one-covariate (age) simple linear
regression for $\widehat{D}(\cdot)$.}
\label{fig:linearCI}
\end{figure}

\begin{figure}[htbp]
\centering
\includegraphics[width=0.45\textwidth]{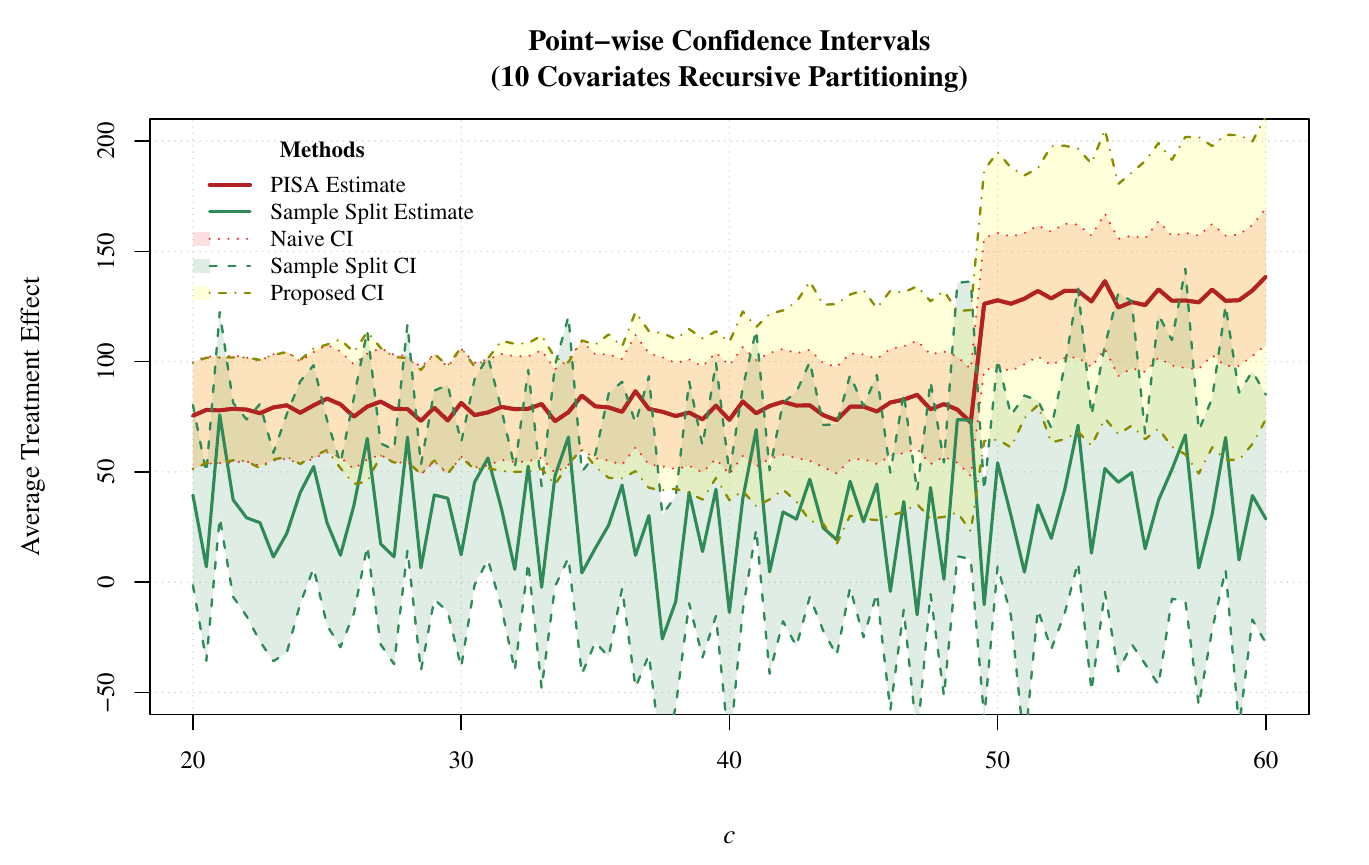}\hspace{0.05\textwidth}%
\includegraphics[width=0.45\textwidth]{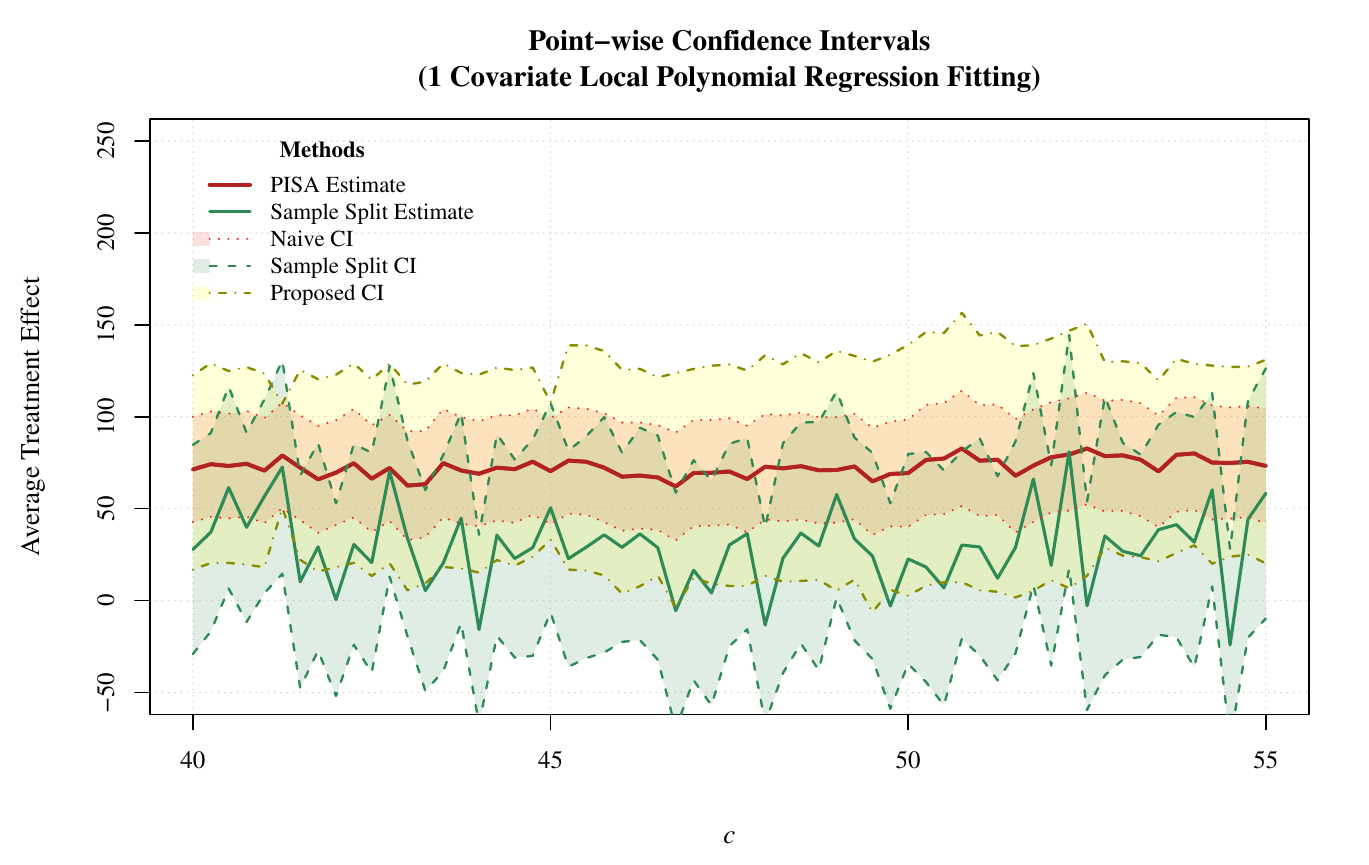}
\caption{Pointwise 95\% confidence intervals for $PISA(c)$. Left
panel: 10-covariate recursive partitioning \citep{su2009subgroup}
for $\widehat{D}(\cdot)$. Right panel: one-covariate (age) local
polynomial regression fitting \citep{cleveland2017local} for
$\widehat{D}(\cdot)$.}
\label{fig:nonparCI}
\end{figure}

Here, we conduct a subgroup analysis of ZDV+ddI versus ZDV+zal based
on the data from the ACTG 175 trial. Specifically, we focus on all
the 1046 patients who receive the treatment, ZDV+ddI, or the control,
ZDV+zal, in the ACTG 175 trial and consider CD4 count at $20\pm 5$
weeks after receiving the therapy as the outcome. To identify
subgroups, we focus on 10 baseline covariates: age (in years),
weight (in kilograms), homosexual activity (yes vs no), history of
intravenous drug use (yes vs no), Karnofsky score (0--100 scale),
history of zidovudine use (receiving zidovudine 30 days prior to
treatment initiation vs no), race (white vs non-white), gender
(female vs male), antiretroviral history (classical vs experienced)
and antiretroviral history stratification (1=`antiretroviral
classical', 2=`$>1$ but $\leq 52$ weeks of prior antiretroviral
therapy' and 3=`$>52$ weeks'). We calculate the estimated PISA
treatment effect curve by Eq.~\eqref{PISA^hat} and implement the
proposed method in Algorithm~\ref{alg-con-moon} with a data-adaptive
choice of the perturbation size $m$ by Algorithm~2 under
both parametric and nonparametric generic machine learning working
model of $\widehat{D}(\cdot)$, and compare the proposed 95\%
confidence intervals with those of the naive method and sample
split, which are in-sample and out-of-sample respectively, as
summarized in Figures~\ref{fig:linearCI} and~\ref{fig:nonparCI}.

Figure~\ref{fig:linearCI} shows that under parametric subgroup
identification, the proposed confidence interval is wider than that
of the naive method in both the 10-covariate and the 1-covariate
(age) cases.  The gap between the two intervals and the instability
of the sample split estimate suggest nonregularity in this study,
and the wider proposed interval is consistent with the known
tendency of the naive method to be overly optimistic under
selection bias.  Notably, when $c$ is around 50, the right panel
shows that the proposed 95\% interval covers 0 while the naive and
sample-split intervals do not, so at the 5\% significance level, the proposed
method would not declare the treatment effect of ZDV+ddI (versus
ZDV+zal) statistically significant for the 1-covariate subgroup
there, while the naive and sample-split methods would.  Given
the potential nonregularity, the proposed finding is the more
credible one.  The estimated PISA curve in Figure~\ref{fig:linearCI}
is also non-monotone, suggesting the linear working model is
misspecified and motivating a nonparametric alternative.

Figure~\ref{fig:nonparCI} shows that under nonparametric subgroup
identification the estimated PISA is more monotone, suggesting that
the nonparametric working model better captures the heterogeneity
in this study.  Under this model neither the naive nor the sample
split method delivers a valid confidence interval, while the
proposed method does.  The jump in the proposed interval around
$c=50$ in the left panel suggests that ZDV+ddI is particularly
preferred in the 10-covariate post-hoc subgroup at $c=50$.

\section{Discussion}
\label{sec:disc}

We have proposed a conditional adaptive perturbation approach for
valid in-sample inference on subgroups identified from the whole
dataset by generic machine learning.  Built on a triply robust
framework linking the rates of subgroup identification and nuisance
estimation, the method addresses selection bias under
nonregularity, accommodates model-free or even black-box working
models, and achieves full efficiency in broad scenarios.  The ACTG
175 analysis illustrates how it can support more replicable
subgroup findings in practice.

More broadly, debiased in-sample evaluation is desired for
data-dependent quantities identified by model-free or black-box
procedures whenever smoothness is potentially lacking.  Analogous
issues arise in machine learning model assessment
\citep{li2020network}: evaluating a metric such as predictive error
rate on the same data used to train a black-box model suffers from
the same kind of selection bias.  Extending our approach to that
setting requires characterizing the bias, stabilizing the
distribution of the in-sample assessment, and linking the rates of
machine learning and nuisance estimation, which we leave to future
work.

\section*{Acknowledgements}

The work was partially supported by grants from Research Grants
Council of the Hong Kong Special Administrative Region, China
(HKUST~26308323 and HKUST~16310125), the Seed fund of the Big Data
for Bio-Intelligence Laboratory (Z0428) and the grant L0438 from the
Hong Kong University of Science and Technology.

\section*{Data Availability Statement}

The ACTG~175 data that support the findings in
Section~\ref{sec:real} of this paper are publicly available as part
of the R package \texttt{speff2trial} on the Comprehensive R Archive
Network (\url{https://cran.r-project.org/package=speff2trial}).
Simulation code reproducing the numerical results in
Section~\ref{sec:simu} is provided as Supplementary Material.

\section*{Supplementary Material}

Supplementary Material A--E and the R code to reproduce the numerical
results referenced in Sections~\ref{sec:setting}, \ref{sec:challenge}, \ref{sec:simu} and~\ref{sec:real} are available online.


\bibliographystyle{apalike}
\bibliography{refs}

\end{document}